\documentclass[conference]{IEEEtran}
\usepackage{times}
\usepackage{xcolor}
\usepackage{xfrac}
\usepackage{soul}
\usepackage[utf8]{inputenc}
\usepackage{amsmath}
\usepackage{tabularx}

\graphicspath{{figs/}} 
\newcommand{\sys}{DeepSigns} 
\usepackage{tabstackengine}
\stackMath
\usepackage{amsfonts,amssymb}
\usepackage[small]{caption}
\usepackage{amsfonts}
\usepackage{multirow}
\usepackage{subcaption}
\usepackage{lipsum}

\begin{document}
\title{Performance Comparison of Contemporary DNN Watermarking Techniques}

\author{
{Huili Chen, Bita Darvish Rouhani, Xinwei Fan, Osman Cihan Kilinc,
and Farinaz Koushanfar}
\\ 
University of California, San Diego  \\
huc044@ucsd.edu, bita@ucsd.edu, x5fan@ucsd.edu, okilinc@ucsd.edu
farinaz@ucsd.edu
}
\maketitle
\begin{abstract}
DNNs shall be considered as the intellectual property (IP) of the model builder due to the impeding cost of designing/training a highly accurate model. Research attempts have been made to protect the authorship of the trained model and prevent IP infringement using DNN watermarking techniques. In this paper, we provide a comprehensive performance comparison of the state-of-the-art DNN watermarking methodologies according to the essential requisites for an effective watermarking technique. We identify the pros and cons of each scheme and provide insights into the underlying rationale. Empirical results corroborate that DeepSigns framework proposed in~\cite{rouhani2018deepsigns} has the best overall performance in terms of the evaluation metrics. Our comparison facilitates the development of pending watermarking approaches and enables the model owner to deploy the watermarking scheme that satisfying her requirements.  
\end{abstract}

\IEEEpeerreviewmaketitle

\vspace{-0.5em}
\section{Introduction}  \label{sec:intro}

Deep neural networks (DNNs) are increasingly commercialized due to their unprecedented performance. Training a DNN is expensive since it requires: (i) availability of sufficient amounts of proprietary data that captures different scenarios in the target application; and (ii) allocation of DL design experts and extensive computational resources to carefully fine-tune of the network topology (i.e., type and number of hidden layers), hyper-parameters (i.e., learning rate, batch size, etc.), and weights. With the growing deployment in various fields, high-performance DNNs shall be considered as the IP of the model owner and need to be protected.

Several papers have proposed leveraging digital watermarking to address the IP protection concern in the deep learning (DL) domain. Existing DNN watermarking techniques can be categorized into two types based on the underlying assumptions. \textit{White-box watermarking} encodes the watermark information (typically a `multi-bit' binary string) in the model internals (e.g., weights, activations) and assumes that these internal details are known to the public. This assumption holds when DNNs are voluntarily shared in the model distribution system. \textit{Black-box watermarking} assumes that the model owner only has API access to the remotely deployed model (send queries and receive the corresponding outputs). The black-box application scenario is common in Machine Learning as a Service (MLaaS) systems~\cite{ribeiro2015mlaas}.

In this paper, we implement the state-of-the-art DNN watermarking methodologies and compare their performance according to a set of essential metrics for an effective watermarking technique. As we empirically corroborate, \sys{} framework proposed in~\cite{rouhani2018deepsigns} has the best overall performance and respects all criteria. \sys{} is also the first and the only watermarking framework that is applicable in both white-box and black-box settings.

\section{Background}   \label{sec:bg}
In this section, we survey the present white-box and black-box DNN watermarking papers. To provide a fair comparison, we deploy the requirements for an effective DNN watermarking methodology as shown in Table~\ref{tab:required}. For details about the definition of each criterion, we refer the readers to the paper~\cite{rouhani2018deepsigns}. We summarize the workflow, advantages, and disadvantages of each watermarking method in Table~\ref{tab:WM_survey}. The quantitative performance comparison of these techniques is given in Section~\ref{sec:eval}.
Potential watermark (WM) removal attacks are discussed in Section~\ref{sec:WM_attacks}.

\begin{table*}
\centering
\caption{Requirements for an effective DNN watermarking scheme~\cite{rouhani2018deepsigns}.}
\label{tab:required}
\vspace{-0.5em}
\scalebox{0.97}{
\begin{tabular}{|l||p{16.2cm}|}
\hline
\multicolumn{1}{|l||}{\textbf{Requirements}}   & \multicolumn{1}{|c|}{\textbf{Description}} \\ \hline \hline
\textbf{Fidelity}     & Accuracy of the target neural network shall not be degraded as a result of watermark embedding. \\ \hline
\textbf{Reliability}  & Watermark extraction shall yield minimal false negatives; WM shall be effectively detected using the pertinent keys. \\ \hline
\textbf{Robustness}   & Embedded watermark shall be resilient against model modifications such as pruning, fine-tuning, or WM overwriting. \\ \hline
\textbf{Integrity }   & Watermark extraction shall yield minimal false alarms (a.k.a., false positives); the watermarked model should be uniquely identified using the pertinent keys. \\ \hline
\textbf{Capacity }    & Watermarking methodology shall be capable of embedding a large amount of information in the target DNN. \\ \hline
\textbf{Efficiency}   & Communication and computational overhead of watermark embedding and extraction shall be negligible. \\ \hline
\textbf{Security }    & The watermark shall be secure against brute-force attacks and leave no tangible footprints in the target neural network; thus, an unauthorized party cannot detect/remove the presence of a watermark. \\ \hline
\end{tabular}}
\end{table*}

\newcolumntype{L}[1]{>{\raggedright\arraybackslash}p{#1}}
\newcolumntype{C}[1]{>{\centering\arraybackslash}p{#1}}
\newcolumntype{R}[1]{>{\raggedleft\arraybackslash}p{#1}}
\begin{table*}
\centering
\caption{Summary of contemporary DNN watermarking methods.}
\label{tab:WM_survey}
\scalebox{1.05}{
\begin{tabular}{|C{1cm}|L{1cm}|L{6cm}|L{3.5cm}|L{3.5cm}|}
\hline
\textbf{Setting} & \multicolumn{1}{c|}{\textbf{Paper}} & \multicolumn{1}{c|}{\textbf{Scheme}} & \multicolumn{1}{c|}{\textbf{Pros}} & \multicolumn{1}{c|}{\textbf{Cons}} \\ \hline
\multirow{2}{*}{\textbf{White-box}} & Uchida et al.~\cite{uchida2017embedding} & Embeds the multi-bit WM in the weights of the selected layer(s) of the target DNN by adding an additive binary cross-entropy loss as the WM regularization term. & Retains the accuracy of the model;  Robust against model compression and fine-tuning attacks. & Vulnerable to WM overwriting attacks. WM is not data-aware. \\
\cline{2-5} 
 & Rouhani et al.~\cite{rouhani2018deepsigns}
 & Embeds the multi-bit WM in the activation maps of the selected layer(s) of the target DNN by incorporating two additional regularization loss terms (GMM center loss and binary cross-entropy loss). & Simultaneous model-aware and data-aware; Robust against WM overwriting, parameter pruning, and model fine-tuning. & WM embedding requires more computation. \\ \hline
\multirow{4}{*}{\textbf{Black-box}} & Merrer et al.~\cite{merrer2017adversarial} & Crafts a set of adversarial samples as the WM key set to alter the decision boundary of the target DNN. The designed WM key image label pairs serve as a zero-bit WM. & Efficient WM embedding and detection. & Accuracy might degrade after WM embedding; Incurs a high false positive rate; Not robust against attacks. \\ \cline{2-5} 
& Yossi et al.~\cite{yossi} & Leverages backdoor images (misclassified) and random labels as the WM key set to embed the zero-bit WM in the pertinent DNN. & Preserve the accuracy of the model; Yields a high detection rate. & Susceptible to WM removal attacks; High overhead to embed the WM. \\ \cline{2-5} 
& Zhang et al.~\cite{zhang2018protecting} & Proposes three key generation algorithms (content-based, unrelated-based, noise-based) to craft the WM key images. & Provides three different types of WM key generation methods.  & Inconsistent performance across different benchmarks. \\ \cline{2-5} 
& Rouhani et al~\cite{rouhani2018deepsigns} & Explores the rarely occupied space within the target DNN to design random image and label pairs as the WM key set. & Yields a high detection rate and a low false alarm rate. & Key generation process incurs higher overhead. \\ \hline
\end{tabular}
}
\end{table*}

\vspace{-0.3em}
\subsection{White-box Watermarking}  \label{sec:white_wm}

White-box WMs have the advantage of larger capacity (carrying more information). However, the capability to convey abundant information comes at the cost of the limited application scenarios since the model internals are required to be publicly available for WM extraction. The Bit Error Rate (BER) between the extracted WM and the ground-truth one needs to be zero to prove the authorship of the queried model. There are two papers providing white-box watermarking techniques and we discuss the mechanism of each approach as follows.  

\noindent \textbf{Uchida et al.} The paper~\cite{uchida2017embedding} presents the first DNN watermarking technique that embeds the owner's watermark information (a binary string) in the weights of the selected layer(s) in the target model. More specifically, the WM is encoded in distribution of the weights by training the DNN with a customized regularization loss that penalizes the difference between the desired WM and the transformation of weights. The WM is later extracted from the queried model by computing the transformation of the marked weights. 

\noindent \textbf{Rouhani et al.} \sys{} proposed in~\cite{rouhani2018deepsigns} is the first generic watermarking framework that is applicable in both white-box and black-box settings. We describe its workflow of white-box watermarking here. DeepSigns assumes a Gaussian Mixture Model (GMM) as the prior probability distribution (pdf) for the activation maps. The WM is embedded in the pdf of activations triggered by the selected WM key images. Two WM-specific regularization loss terms are incorporated during DNN training to align the activations and encode the WM information. In the WM extraction stage, the WM key images are passed through the queried model and trigger the marked activations. The WM is recovered from the transformation of the resulting activations and the BER is computed.

\subsection{Black-box Watermarking} \label{sec:black_wm}

Current black-box watermarking methods~\cite{merrer2017adversarial, zhang2018protecting, rouhani2018deepsigns, yossi} target at `zero-bit' watermarking, which is only concerned about the existence of the WM. A set of WM key pairs is generated to strategically alter the decision boundary of the target model. The presence of the WM is determined by querying the model with the WM key images and thresholding the corresponding accuracy. Black-box WMs are more practical in the real-world settings due to the relaxed assumption (requiring API access instead of model internals) but possesses limited capacity. We discuss the working mechanisms of existing black-box watermarking techniques as follows.

\noindent \textbf{Merrer et al.}~\cite{merrer2017adversarial} proposes to craft adversarial samples as the WM key set for WM embedding.
When the adversarial attack succeeds, the corresponding sample are referred to as `true adversaries' that carries the WM information. When the attack fails, the generated images are referred to as `false adversaries' that are used to preserve the accuracy of the model on legitimate data. The combination of true and false adversaries forms the complete WM key set. In the WM detection phase, the model is queried by the WM key images and a statistical null-hypothesis testing is performed assuming a binomial distribution. If the number of mismatches between the model's response and the WM key labels is smaller than the threshold, the WM is decided to be existent in the model.

\noindent \textbf{Yossi et al.}~\cite{yossi} suggests to use the backdoors (images that are misclassified by the model) as the WM key images. The corresponding key label is randomly sampled from all classes excluding the ground-truth label and the original predicted one. The WM is detected by comparing the accuracy on the WM trigger set with the threshold determined by binomial distribution. Furthermore,~\cite{yossi} suggests to deploy the commitment scheme for constructing a publicly verifiable protocol.

\noindent \textbf{Zhang et al.}~\cite{zhang2018protecting} presents three different WM key generation methods that leverage unrelated images in another dataset, the training image superimposed with additional meaning content, and random images, respectively. The WM keys are then used to fine-tune the pre-trained model. During the WM detection stage, the owner sends the WM key images to the remote DNN service provider and thresholds the classification accuracy to make the Boolean decision.  

\noindent \textbf{Rouhani et al.}~\cite{rouhani2018deepsigns} first generates an initial WM key set that consists of random image and random labels pairs. The initial key set is used for two purposes: (i) identifying the key pairs that are misclassified by the original unmarked model; and (ii) fine-tuning the target model for WM embedding. The final WM key set is the intersection of the keys that are correctly predicted by the marked model and falsely predicted by the unmarked model. To detect the watermark,~\cite{rouhani2018deepsigns} assumes a multinomial distribution for the output prediction and performs the null-hypothesis  testing with the final WM key set.  
 
\subsection{Attack Model} \label{sec:WM_attacks}

To validate the robustness of a potential DL watermarking approach, one should evaluate the robustness of the proposed methodology against (at least) three types of contemporary attacks: (i) \textbf{model fine-tuning}. This attack involves re-training of the original model to alter the model parameters and find a new local minimum while preserving the accuracy. (ii) \textbf{model pruning}. Model pruning is a common approach for efficient DNN execution, particularly on embedded devices. We identify model pruning as another attack approach that might affect the watermark extraction/detection. (iii) \textbf{watermark overwriting}. A third-party user who is aware of the methodology used for DNN watermarking (but does not know the owner's private WM keys) may try to embed a new watermark in the model and overwrite the original one. An overwriting attack intends to insert an additional watermark in the model and render the original one undetectable. A watermarking methodology should be robust against fine-tuning, pruning, and overwriting for effective IP protection.

\section{Evaluations}  \label{sec:eval}

In this section, we provide a quantitative comparison of the papers summarized in Table~\ref{tab:WM_survey}. We implement the white-box weights watermarking~\cite{uchida2017embedding} using their open-sourced code in~\cite{uchida_code}. All the other watermarking techniques are implemented based on the work flow and experimental setup described in the original papers. It is worth noting that~\cite{zhang2018protecting} does not provide the decision threshold for WM detection. We use as $80\%$ accuracy on the WM key set as the WM detection threshold in our experiments to provide a fair comparison with other watermarking methods. We explicitly compare the watermarking performance regarding each metric discussed in Table~\ref{tab:required} as follows.

\vspace{-0.5em}
\subsection{Fidelity} \label{sec:fidelitry_eval}
\vspace{-0.3em}
To assess the fidelity of the watermarking techniques, we compare the test accuracy of the target model before and after WM embedding. The results in the white-box and black-box setting are shown in Table~\ref{tab:white_acc} and Table~\ref{tab:black_acc}, respectively. The averaged test accuracy of the marked model using three different WM key generation algorithms in~\cite{zhang2018protecting} is reported in the last row of Table~\ref{tab:black_acc}. One can be see that both~\cite{uchida2017embedding} and~\cite{rouhani2018deepsigns} white-box watermarking methodologies are able to preserve the accuracy of the original model (shown in the parenthesis) due to the simultaneous optimization of the conventional cross-entropy loss together with the WM-specific regularization loss.

As for the black-box watermarking approaches,~\cite{merrer2017adversarial} suffers from accuracy degradation on two CIFAR-10 benchmarks while the other methods preserve the same level of accuracy compared to the baseline across various benchmarks. The reason is that~\cite{merrer2017adversarial} only uses $K$ crafted adversarial samples ($K=20$ in the experiment) to embed the WM, which does not capture sufficient coverage of the data in the target application. The other three watermarking methods use a mixture of the original training data (or a subset of it) and the designed WM key images, thus do not induce a significant accuracy drop.

\vspace{-1em}
\begin{table}[ht!]
\centering
\caption{Fidelity evaluation of two white-box watermarking schemes. The test accuracy after embedding a 4-bit WM and the baseline accuracy (in parenthesis) are shown.}
\label{tab:white_acc}
\vspace{-0.3em}
\scalebox{0.98}{
\begin{tabular}{|l|l|l|l|}
\hline
\multirow{2}{*}{\textbf{White-box WM}} & \multicolumn{3}{c|}{\textbf{Benchmarks}} \\ \cline{2-4} 
 & MNIST-MLP & CIFAR10-CNN & CIFAR10-WRN \\ \hline
Uchida~\cite{uchida2017embedding} & 98.34\% (98.54\%) & 80.06\% (78.47\%) & 91.63\% (91.42\%) \\ \hline
Rouhani~\cite{rouhani2018deepsigns}  & 98.13\% (98.54\%) & 80.70\% (78.47\%) & 92.02\% (91.42\%) \\ \hline
\end{tabular}
}
\end{table}

\vspace{-1em}
\begin{table}[ht!]
\centering
\caption{Fidelity evaluation of four black-box watermarking methods. The WM key length is set to 20. The test accuracy before (in parenthesis) and after WM embedding is compared.}
\label{tab:black_acc}
\vspace{-0.3em}
\scalebox{0.8}{
\begin{tabular}{|l|c|c|c|c|}
\hline
\multirow{2}{*}{\textbf{Black-box WM}} & \multicolumn{4}{c|}{\textbf{Benchmarks}} \\ \cline{2-5} 
 & MNIST-MLP & CIFAR10-CNN & CIFAR10-WRN & ImageNet-ResNet50 \\ \hline
Merrer~\cite{merrer2017adversarial} & 98.24\%(98.54\%) & 77.03\%(80.60\%) & 77.0\%(91.42\%) & 74.73\%(74.73\%) \\ \hline
Rouhani~\cite{rouhani2018deepsigns} & 98.61\% (98.54\%) & 81.48\% (78.47\%) & 92.03\% (91.42\%) & 73.83\% (74.73\%) \\ \hline
Yossi~\cite{yossi} & 97.48\%(98.54) & 80.18\%(80.60\%) & 91.36\%(91.42\%) & 74.73\%(74.73\%) \\ \hline
Zhang~\cite{zhang2018protecting} & 98.51\% (98.54\%) & 77.53 (78.47\%) & 91.65\% (91.42\%) & 74.09\% (74.73\%) \\ \hline
\end{tabular}
}
\end{table}

\vspace{-0.3em}
\subsection{Robustness} 
In the following of this section, we compare the robustness of the embedded WM against three types of WM removal attacks discussed in Section~\ref{sec:WM_attacks}. 

\begin{figure}[ht!]
\centering
  \includegraphics[width=0.92 \columnwidth]{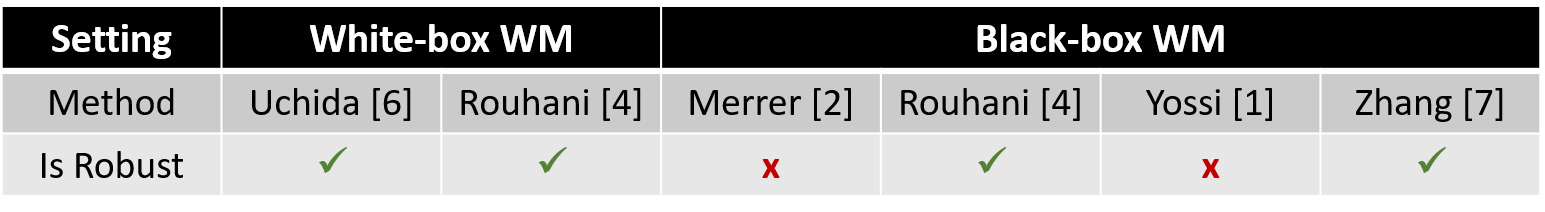} 
\caption{\label{fig:ft_attack} Robustness comparison of watermarking methods against model fine-tuning attacks.}
\end{figure}

\noindent \textbf{(i) Model Fine-tuning Attack.} This attack involves retraining the watermarked model with the conventional cross-entropy loss on the original training dataset. In the white-box setting, a robust WM shall be able to yield zero BER after the weights/activations are changed during model fine-tuning. In the black-box setting, the prediction accuracy of the queried model on the WM key set shall be larger than the decision threshold for the WM to be robust. The results in both scenarios are summarized in Table~\ref{fig:ft_attack}. A watermarking method is robust if the embedded WM can withstand the fine-tuning attack across all benchmarks. We use a single column to denote its performance for simplicity since all three WM key generation algorithms are robust.

\begin{figure}[ht!]
\centering
  \includegraphics[width=0.92\columnwidth]{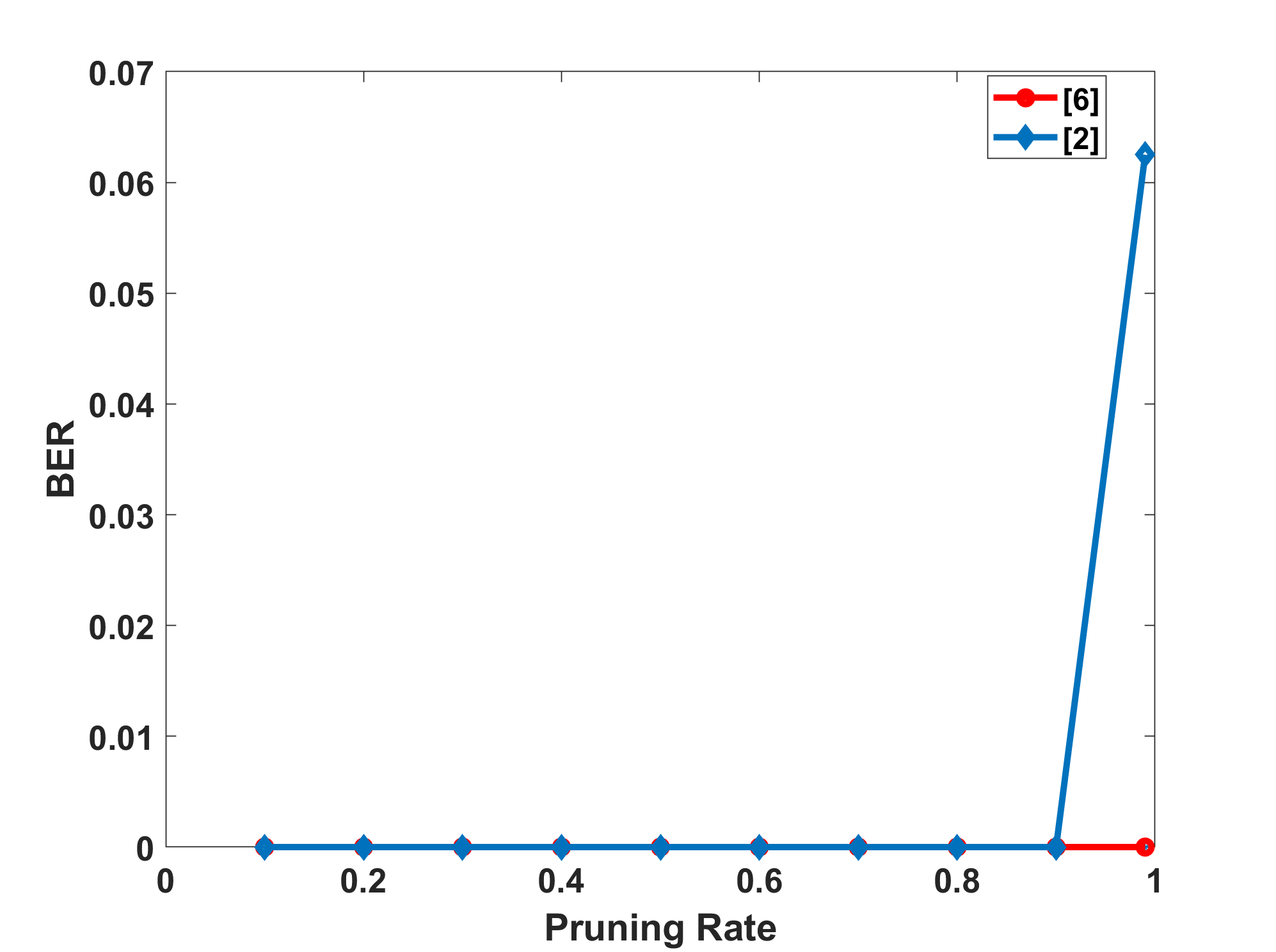} 
\caption{\label{fig:white_prune_cifar10wrn} Robustness comparison of two white-box watermarking methods against parameter pruning attacks. A binary string of length 128 is embedded in the second last dense layer of the WRN model.}
\end{figure}

\noindent \textbf{(ii) Parameter Pruning Attack.} We prune the model by setting the weights with small magnitudes to zeros. We first compare the robust of two white-box watermarking frameworks on three benchmarks shown in Table~\ref{tab:white_acc}. Since the relative performance of~\cite{uchida2017embedding} and~\cite{rouhani2018deepsigns} is the same across different benchmarks, we only visualize the results on CIFAR10-WRN benchmark in Figure~\ref{fig:white_prune_cifar10wrn}. one can see that: (i). both~\cite{uchida2017embedding} and~\cite{rouhani2018deepsigns} can withstand a wide range of parameter pruning; (ii). directly embedding the WM information in weights~(\cite{uchida2017embedding}) is slightly more robust against pruning attacks compared to activation-based WM embedding~(\cite{rouhani2018deepsigns}).

\vspace{-1em}

\begin{figure}[ht!]
\centering
  \includegraphics[width=0.92\columnwidth]{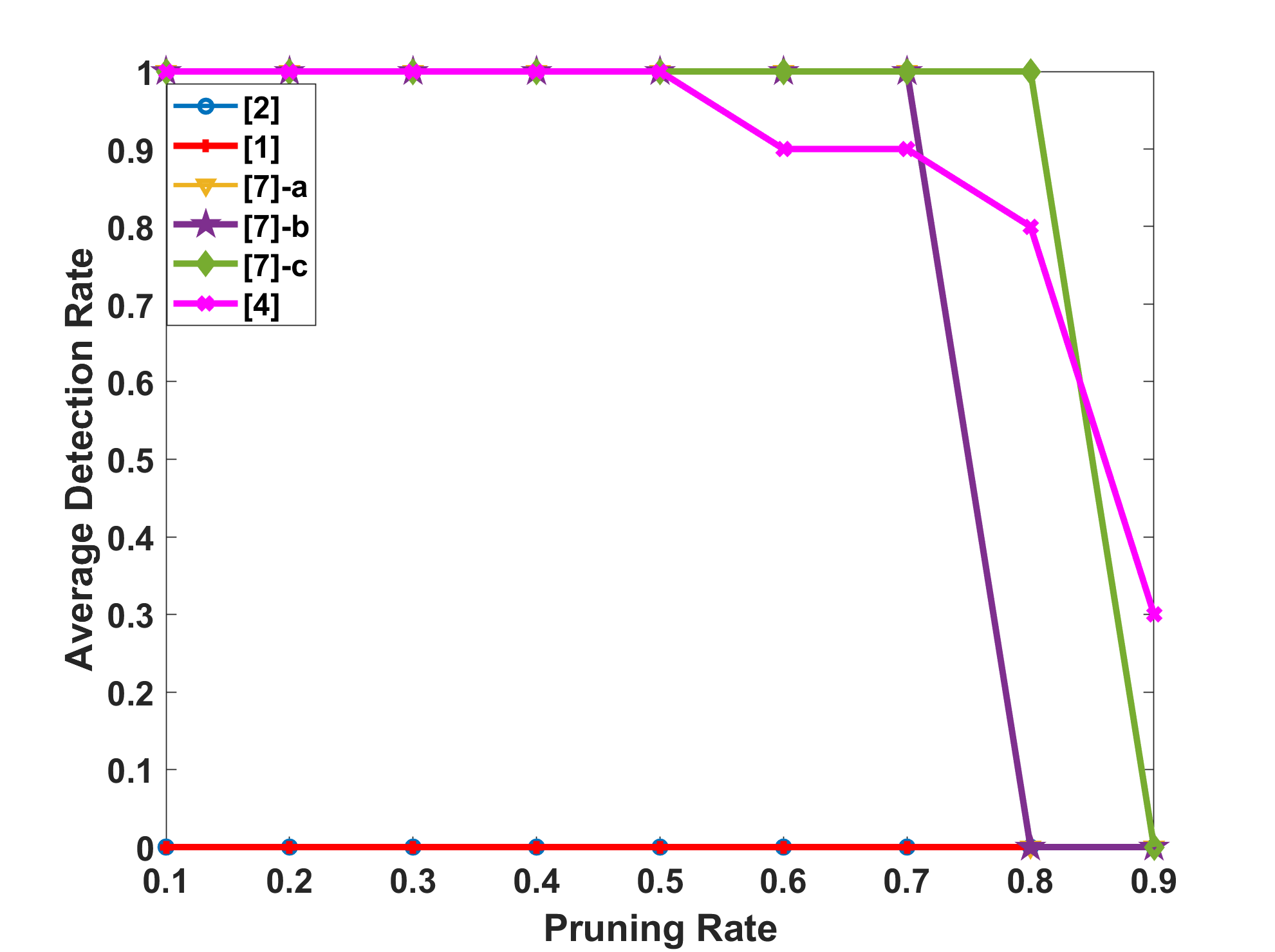} 
\caption{\label{fig:black_prune_cifar10cnn} Robustness comparison of four black-box watermarking methods against parameter pruning attacks on CIFAR10-CNN benchmark. Note that~\cite{merrer2017adversarial},~\cite{yossi}, and the content-based WM in~\cite{zhang2018protecting} has zero detection rate on this benchmark and their corresponding lines completely overlap in the plot. }
\end{figure}

\begin{figure}[ht!]
\centering
  \includegraphics[width=0.92\columnwidth]{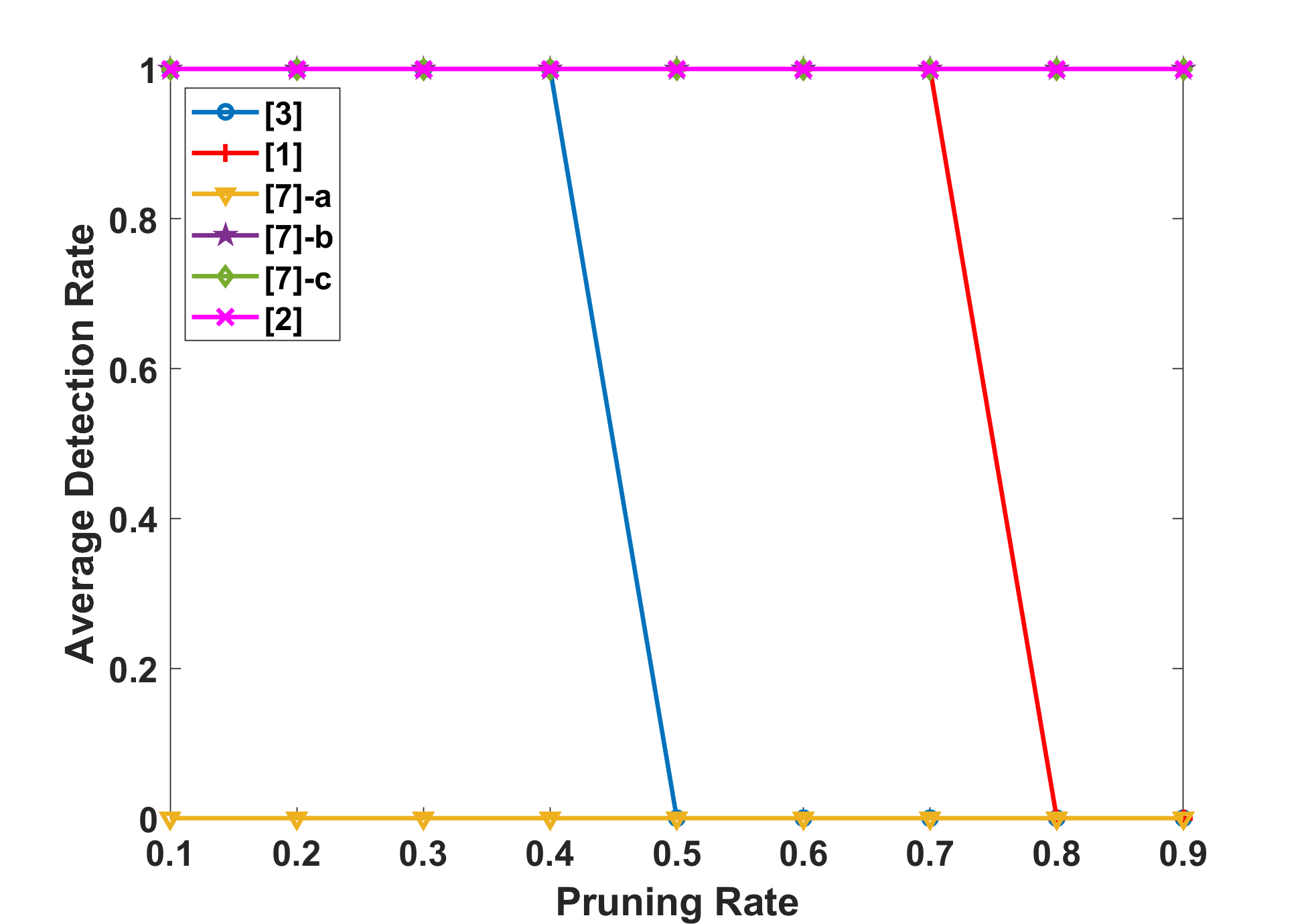} 
\caption{\label{fig:black_prune_resnet} Average WM detection rates of black-box watermarking methods under different pruning rates on the ImageNet-ResNet benchmark. Note that all three WM key generation methods in~\cite{zhang2018protecting} are vulnerable to parameter pruning, yielding zero detection rate on this benchmark.}
\end{figure}

We also compare the robustness of four black-box watermarking methods against parameter pruning attacks. Since the performance of~\cite{merrer2017adversarial} and~\cite{yossi} is not uniform across different benchmarks, we show the results on CIFAR10-WRN and ImageNet-ResNet in Figure~\ref{fig:black_prune_cifar10cnn} and Figure~\ref{fig:black_prune_resnet}. The WM key length is set to $K=20$. The content-based, unrelated-based, noise-based WM key generation methods in~\cite{zhang2018protecting} are denoted as `a', `b', `c' in the figures, respectively. It can be seen that~\cite{rouhani2018deepsigns} and~\cite{zhang2018protecting} has consistent high WM detection rate on middle and large scale datasets when the model is pruned for a wide range of values. The noise-based WM key generation method in~\cite{zhang2018protecting} is more robust against parameter pruning attacks compared to the content-based and unrelated-based methods.~\cite{merrer2017adversarial} and~\cite{yossi} yield lower detection rates against parameter pruning attacks on CIFAR10-CNN compared to ImageNet-ResNet benchmark, which makes the watermarking methods not desirable for practical usage.

\vspace{-1.3em}

\vspace{-2em}

\vspace{-0.8em}
\noindent \textbf{(iii) Watermark Overwriting Attack.}

Table~\ref{tab:compuchida} provides a side-by-side robustness comparison between~\cite{rouhani2018deepsigns} and~\cite{uchida2017embedding} white-box watermarking methodologies for different dimensionality ratio (defined as the ratio of the length of the attacker's WM signature to the size of weights or activations). As can be seen from the table, embedding WM in the dynamic activation maps suggested by \sys{} is more robust against WM overwriting attack compared to embedding WM in the weights as proposed in~\cite{uchida2017embedding}.

\begin{table}[ht!]
\centering
\vspace{-1em}
\caption{Robustness comparison against WM overwriting attacks on CIFAR10-WRN benchmark. The WM information embedded by~\cite{rouhani2018deepsigns} can withstand overwriting attacks for a wide of range of $\frac{N}{M}$ ratio compared to~\cite{uchida2017embedding}.}
\label{tab:compuchida}
\resizebox{0.35\textwidth}{!}{%
\begin{tabular}{|c||c|c|}
\hline
\multirow{2}{*}{\textbf{N to M Ratio}} & \multicolumn{2}{c|}{\textbf{Bit Error Rate (BER)}} \\ \cline{2-3} 
 & \textbf{Uchida et al.~\cite{uchida2017embedding}} & \textbf{Rouhani et al.~\cite{rouhani2018deepsigns}} \\ \hline \hline
1 & 0.309 & 0 \\ 
2 & 0.41 & 0 \\ 
3 & 0.511 & 0 \\ 
4 & 0.527 & 0 \\ \hline
\end{tabular}%
}
\vspace{-0.5em}
\end{table}

\begin{figure}[ht!]
\centering
  \includegraphics[width=0.86\columnwidth]{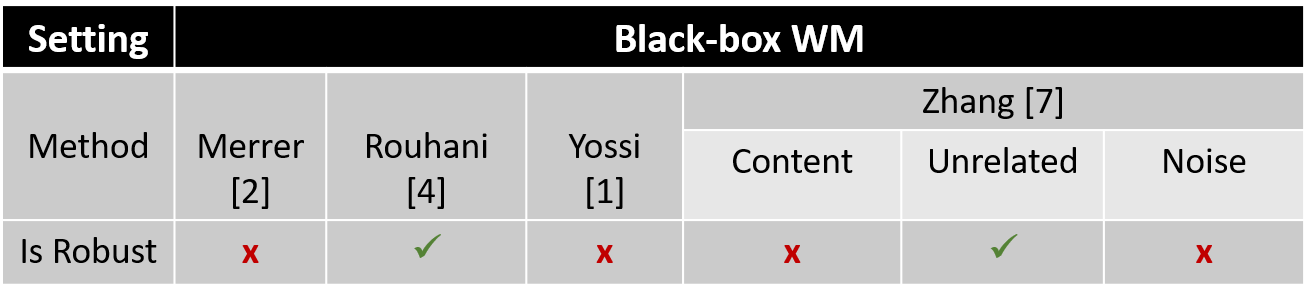} 
\caption{\label{fig:black_overwrite} Robustness comparison of black-box watermarking methods against WM overwriting attacks.}
\end{figure}

Figure~\ref{fig:black_overwrite} illustrates the robustness comparison results of black-box watermarking methods against WM overwriting attacks. The approach is marked as `robust' only when it withstands the overwriting attack across all benchmarks shown in Table~\ref{tab:black_acc}. It can be seen that~\cite{rouhani2018deepsigns} and the unrelated WM in~\cite{zhang2018protecting} are robust against WM overwriting attacks.

\vspace{-1.2em}
\subsection{Integrity}
Integrity requires that the authorship of an unmarked model will not be falsely claimed by the watermarking methodology. In the white-box setting, if the owner tries to extract a WM from an unmarked model, the computed BER shall be a non-zero to satisfy the integrity requirement. We follow the WM extraction procedure in paper~\cite{uchida2017embedding} and~\cite{rouhani2018deepsigns} to evaluate their integrity. The resulting BER has a large value (around $0.5$) in both cases, suggesting these two white-box watermarking methods satisfy the integrity criterion. 

\begin{figure}[ht!]
\centering
  \includegraphics[width=0.86\columnwidth]{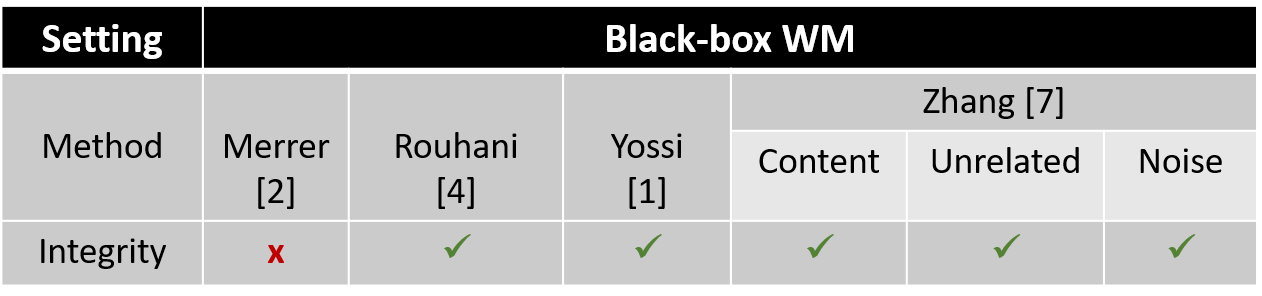} 
\caption{\label{fig:black_integrity} Integrity comparison of black-box watermarking methods.}
\end{figure}

In the black-box setting, integrity requires that the existence of the WM will not be falsely detected by the watermarking scheme. We assess the integrity of each black-box watermarking method in Table~\ref{tab:black_acc} as follows. Three unmarked models with the same and different topologies as the watermarked variant are evaluated. The watermarking scheme is decided to be `integrate' when the WM is not detected in all six models across various benchmark. Figure~\ref{fig:black_integrity} shows the integrity comparison results in the black-box scenario. Only the watermarking method in~\cite{merrer2017adversarial} violates the integrity requirement (yields a high false alarm rate) since the crafted WM key images are adversarial samples that are transferable to other models which might not be watermarked.


\section{Conclusion}  \label{sec:concl}

We provide a comprehensive qualitative and quantitative comparison of the state-of-the-art DNN watermarking methodologies in both white-box and black-box settings. The advantages and disadvantages of each evaluated watermarking scheme are identified. Experimental results corroborate that \sys{} framework proposed in~\cite{rouhani2018deepsigns} outperforms the other watermarking techniques by designing a robust WM that is applicable in both application scenarios. Our side-by-side performance comparison helps the research communities to develop more advanced watermarking methodologies. Practitioners can also use the comparison results as a reference to determine the appropriate watermarking framework that satisfies their performance requirements.

\bibliographystyle{IEEEtranS}
\bibliography{ref}

\begin{thebibliography}{1}
\providecommand{\url}[1]{#1}
\csname url@samestyle\endcsname
\providecommand{\newblock}{\relax}
\providecommand{\bibinfo}[2]{#2}
\providecommand{\BIBentrySTDinterwordspacing}{\spaceskip=0pt\relax}
\providecommand{\BIBentryALTinterwordstretchfactor}{4}
\providecommand{\BIBentryALTinterwordspacing}{\spaceskip=\fontdimen2\font plus
\BIBentryALTinterwordstretchfactor\fontdimen3\font minus
  \fontdimen4\font\relax}
\providecommand{\BIBforeignlanguage}[2]{{%
\expandafter\ifx\csname l@#1\endcsname\relax
\typeout{** WARNING: IEEEtranS.bst: No hyphenation pattern has been}%
\typeout{** loaded for the language `#1'. Using the pattern for}%
\typeout{** the default language instead.}%
\else
\language=\csname l@#1\endcsname
\fi
#2}}
\providecommand{\BIBdecl}{\relax}
\BIBdecl

\bibitem{yossi}
Y.~Adi, C.~Baum, M.~Cisse, B.~Pinkas, and J.~Keshet, ``Turning your weakness
  into a strength: Watermarking deep neural networks by backdooring,''
  \emph{Usenix Security Symposium}, 2018.

\bibitem{merrer2017adversarial}
E.~L. Merrer, P.~Perez, and G.~Tr{\'e}dan, ``Adversarial frontier stitching for
  remote neural network watermarking,'' \emph{arXiv preprint arXiv:1711.01894},
  2017.

\bibitem{ribeiro2015mlaas}
M.~Ribeiro, K.~Grolinger, and M.~A. Capretz, ``Mlaas: Machine learning as a
  service,'' in \emph{IEEE 14th International Conference on Machine Learning
  and Applications (ICMLA)}, 2015.

\bibitem{rouhani2018deepsigns}
B.~D. Rouhani, H.~Chen, and F.~Koushanfar, ``Deepsigns: A generic watermarking
  framework for ip protection of deep learning models,'' \emph{arXiv preprint
  arXiv:1804.00750}, 2018.

\bibitem{uchida_code}
Y.~Uchida \emph{et~al.}, ``Embedding watermarks into deep neural networks,''
  \url{https://github.com/yu4u/dnn-watermark}, 2017.

\bibitem{uchida2017embedding}
Y.~Uchida, Y.~Nagai, S.~Sakazawa, and S.~Satoh, ``Embedding watermarks into
  deep neural networks,'' in \emph{Proceedings of the ACM on International
  Conference on Multimedia Retrieval}, 2017.

\bibitem{zhang2018protecting}
J.~Zhang, Z.~Gu, J.~Jang, H.~Wu, M.~P. Stoecklin, H.~Huang, and I.~Molloy,
  ``Protecting intellectual property of deep neural networks with
  watermarking,'' in \emph{Proceedings of the 2018 on Asia Conference on
  Computer and Communications Security}.\hskip 1em plus 0.5em minus 0.4em\relax
  ACM, 2018, pp. 159--172.

\end{thebibliography}
\end{document}